\documentclass[sigconf]{acmart}

\def\BibTeX{{\rm B\kern-.05em{\sc i\kern-.025em b}\kern-.08emT\kern-.1667em\lower.7ex\hbox{E}\kern-.125emX}}
    
\copyrightyear{2018}
\acmYear{2018}
\setcopyright{acmlicensed}
\acmConference[Woodstock '18]{Woodstock '18: ACM Symposium on Neural Gaze Detection}{June 03--05, 2018}{SigKDD, NY}
\acmBooktitle{Woodstock '18: ACM Symposium on Neural Gaze Detection, June 03--05, 2018, Woodstock, NY}
\acmPrice{15.00}
\acmDOI{10.1145/1122445.1122456}
\acmISBN{978-1-4503-9999-9/18/06}

\usepackage{times}
\usepackage{xcolor}
\usepackage{soul}
\usepackage{graphicx}
\usepackage[utf8]{inputenc}
\usepackage{algorithm}
\usepackage[noend]{algpseudocode}
\usepackage{subcaption}
\usepackage{tabulary} 
\usepackage{colortbl,booktabs}
\usepackage{threeparttable}  
\usepackage{diagbox}
\usepackage{multirow}
\usepackage{booktabs}
\usepackage{makecell}

\DeclareMathOperator*{\E}{\mathbb{E}}

\begin{document}

\title{Sequential Evaluation and Generation Framework for Combinatorial Recommender System}


\author{Fan Wang}
\affiliation{
  \institution{Baidu Inc.}
  \city{Shenzhen}
  \country{China}}
\email{wang.fan@baidu.com}
\authornote{Both authors contributed equally to this research.}

\author{Xiaomin Fang}
\affiliation{
  \institution{Baidu Inc.}
  \city{Shenzhen}
  \country{China}}
\email{fangxiaomin01@baidu.com}
\authornotemark[1]

\author{Lihang Liu}
\affiliation{
  \institution{Baidu Inc.}
  \city{Shenzhen}
  \country{China}}
\email{liulihang@baidu.com}

\author{Yaxue Chen}
\affiliation{
  \institution{Baidu Inc.}
  \city{Shenzhen}
  \country{China}}
\email{chenyaxue@baidu.com}

\author{Jiucheng Tao}
\affiliation{
  \institution{Baidu Inc.}
  \city{Shenzhen}
  \country{China}}
\email{taojiucheng@baidu.com}

\author{Zhiming Peng}
\affiliation{
  \institution{Baidu Inc.}
  \city{Beijing}
  \country{China}}
\email{pengzhiming01@baidu.com}

\author{Cihang Jin}
\affiliation{
  \institution{Baidu Inc.}
  \city{Beijing}
  \country{China}}
\email{jincihang@baidu.com}

\author{Hao Tian}
\affiliation{
  \institution{Baidu Research}
  \city{Sunnyvale}
  \country{United States}}
\email{tianhao@baidu.com}
\renewcommand{\shortauthors}{Wang and Fang, et al.}

\begin{abstract}
In the combinatorial recommender systems, multiple items are fed to the user at one time in the result page, where the correlations among the items have impact on the user behavior. In this work, we model the combinatorial recommendation as the problem of generating a sequence(ordered list) of items from a candidate set, with the target of maximizing the expected overall utility(e.g. total clicks) of the sequence. Toward solving this problem, we propose the Evaluation-Generation framework. On the one hand of this framework, an evaluation model is trained to evaluate the expected overall utility, by fully considering the user, item information and the correlations among the co-exposed items. On the other hand, generation policies based on heuristic searching or reinforcement learning are devised to generate potential high-quality sequences, from which the evaluation model select one to expose. We propose effective model architectures and learning metrics under this framework. We also offer series of offline tests to thoroughly investigate the performance of the proposed framework, as supplements to the online experiments. Our results show obvious increase in performance compared with the previous solutions.
\end{abstract}


\maketitle

\section{Introduction}

Recommender Systems(RS) attracts a lot of attention with the booming of information on the Internet. Typical RS algorithms include collaborative methods (\cite{sarwar2001item}, \cite{koren2009matrix}, \cite{he2016fast}), content based methods and hybrid methods of the two(\cite{adomavicius2015context}). Among the recent works, several topics have drawn more attentions: Context-Aware Recommendation(\cite{adomavicius2015context}) seeks to utilize the information of scenes where the users are fed; Time-Aware RS(\cite{campos2014time}) focuses on the evolution of the user interest; Diversified ranking(\cite{wilhelm2018practical}) tries to address the correlation among co-exposed items, i.e. the intra-list correlations.

It's important to consider the intra-list correlations in many realistic RS in order for better user experience. However, the research on intra-list correlations usually collapses to diversity (\cite{wilhelm2018practical}) in many works. Yet we think that the investigation in intra-list correlation is far from enough in several aspects: Firstly, the evaluation of diversity misses a gold standard. Though there have been various metrics such as \textbf{Coverage} (\cite{ziegler2005improving}), \textbf{Intra-List Similarity} (\cite{shani2011evaluating}) etc, those evaluation metrics are typically subjective, and not directly related to the true user experience. Secondly, strong propositions have been imposed on the formulation of the correlation in the previous work. Algorithms such as determinant point process (\cite{wilhelm2018practical}) and sub-modular ranking rely on handcrafted kernels or functions, which cannot effectively capture all possible correlation forms. Thirdly, the traditional step-wise greedy ranking method typically neglects the loss of the local optimum. If we consider the sub-modular ranking, the lower bound of the ratio of the overall utility comparing the greedy choice to the global optimum is $(1-1/e)$(\cite{yue2011linear}). Yet, the loss of the local optimum is totally unclear when going beyond the sub-modularity hypothesis.

In this paper, we propose to optimize the overall utility of a sequence, with the preposition that the diversity and the other intra-list correlations need to be responsible for this utility. To do so, on the one hand, we build neural architectures to encode the list to predict the utility of the list. We call this part the \textbf{Evaluator}. On the other hand, we try to build ranking policies to generate recommendation lists from a candidate set of items, which we call the \textbf{Generator}. We use sequence decoder as the ranking policy, targeting at generating a sequence(a recommendation list) with as high overall utility as possible. This target can be typically realized by heuristic searching or reinforcement learning. In addition, we use the Generator to generate multiple potential high-quality lists, from which the Evaluator further selects the superior one to achieve higher performance. 

Offline evaluation of combinatorial RS is challenging, as static data can not be used to evaluate such system. Previous work either use subject metrics(\cite{feng2018greedy},\cite{ziegler2005improving}) or simualtors(\cite{yue2011linear}), while the simulator itself is typically not validated. In this work, we adopt progressive evaluation steps to validate the Evaluator and the Generator respectively. The validity of the proposed framework is supported by both offline analysis and online experiments.

There are several contributions in this work:
\begin{itemize}
\item We provide a thorough investigation of intra-list correlations in a realistic recommender system. We propose model architectures that capture those correlations in more effective way compared with the previous methods.
\item We propose practical offline evaluation metrics for combinatorial RS, which is consistent with the online experiments.
\item The proposed recommender frameworks is fully launched in an online system with hundreds of millions of daily active users.

\end{itemize}

\section{Related Works}

Diversity has been frequently investigated in the area of intra-list correlations. Recent works on diversified ranking include the submodularity (\cite{azar2011ranking},\cite{yan2017graph}), graph based methods (\cite{zhu2007improving}), and Determinantal Point Process (DPP) (\cite{borodin2009determinantal}, \cite{kulesza2012determinantal}, \cite{wilhelm2018practical}). The diversified ranking with predefined submodular functions typically supposes that diversity are homogeneous on different topics, and independent of the user. DPP and submodular ranking also suppose that co-exposed items always have a negative impact on the possible click of the others. In contrast to those propositions, realistic RS show cases that violate those rules. Our statistics reveal that some related contents are prone to be clicked together. Except for combination, phenomenons relating the user feedback to display positions have also been widely studied, e.g., click models in Information Retrieval(IR) systems, such as the Cascade Click Model(\cite{craswell2008experimental}) and the Dynamic Bayesian Network(\cite{chapelle2009dynamic}). It is found that the position bias is not only related to the user's personal habit, but also related to the layout design etc(\cite{chuklin2015click}). Thus click models often need to be considered case-by-case. More complex phenomenon has also been discovered, such as \textbf{Serpentining}\cite{zhao2017toward}, which found that the users prefer discontinuous clicks when browsing the list. It is recommended that high quality items should be scattered over the entire list instead of clustered on the top positions. In contrast to those discoveries, few the previous works have studied the intra-list correlation and the position bias together. 

Our framework draws ideas from model-based reinforcement learning(\cite{doll2012ubiquity}), where the transition probability and reward function is directly approximated through modeling the environment(here we have the Evaluator), and heuristic searching and other planning techniques are used to search for optimistic trajectories(just like the Generator). We also borrow ideas from applying model-free Reinforcement Learning for effective planning and searching. Various architectures and learning metrics has been proposed for similar problem, e.g. Vinyals et al. applied pointer network for universal combinatorial optimization(CO) problem(\cite{vinyals2015pointer}). Bello et al. further extended pointer-network by using policy gradients for learning(\cite{bello2016neural}). Dai et al. applied Q-Learning to CO problems on graphs(\cite{dai2017learning}). Those works are focused on general optimization problems such as CO, while extension to RS requires further investigation.

Works on addressing different kinds of long-term rewards in RS have also been reported, too. Feng et al. applied policy gradient and Monte Carlo Tree Search(MCTS) to optimize the $\alpha$-NDCG in diversified ranking for the global optimum(\cite{feng2018greedy}). Other works pursued long-term rewards in inter-list recommendations (\cite{zhao2018deep}, \cite{zheng2018drn}). Though Zhao et al. also proposed treating a page of recommendation list as a whole(\cite{zhao2018deep}), the intra-list correlations are not sufficiently analyzed. Our work is fundamentally different in more thoroughly investigation of the intra-list correlations. Though inter-list correlations are also important, we regard it as an independent problem that is not studied in this paper.


\section{Methodology}

\subsection{Problem Setup}

We formulate the combinatorial RS as follow: The environment exposes the user profile $u$ and a candidate set $\textbf{c} = \{c_1, c_2, ..., c_N\}$ to the RS, where $N$ is the cardinality of the candidate set and $c_i$ denotes the $i$-th item. The system picks a recommendation sequence $\textbf{a} = [c_{a_1}, c_{a_2},..., c_{a_K}]$ where $a_j \in [1,N]$ and $N \geq K$. $\textbf{a}$ is exposed to the user, after which the user returns the feedback of $r(u,\textbf{a})$. Furthermore, we denote $\textbf{a}_j^- = [c_{a_1}, c_{a_2}, ... c_{a_{j-1}}]$ as the preceding recommendations above the $j$-th position, and $\textbf{a}_j^+ = [c_{a_{j+1}}, c_{a_{j+2}}, ... c_{a_{K}}]$ as the recommendations that follows. We define the final objective as maximizing the expected overall utility of the sequence $\textbf{a}$, which is written as $\E[r(u,\textbf{a})]$, where we use $\E_{X}[]$ to represent the expectation over variable $X$, $\E[]$ simply represents the expectation over repeated experiments.

\begin{equation}
\label{Eq:SRSTarget}
\textbf{a}_{opt} = argmax_{\textbf{a}} \E[r(u, \textbf{a})]
\end{equation}

Typically, the overall utility $r(u, \textbf{a})$ is the summation of utility(rewards) $r_j(u, \textbf{a})$ (e.g., clicks) over all positions $j \in [1, K]$ in the list, i.e.  $r(u,\textbf{a}) = \sum_{j=1}^K r_j(u,\textbf{a})$. We use $\textbf{r}(u, \textbf{a}) = [r_1, ..., r_K]$ to denote the vector of rewards. Except for clicks, other feedbacks might be avaible in RS, such as dwelling time(either item-wise or dwelling time over the whole list) and request for subsequent recommendation. Some of the rewards can not be assigned to each item, while others can be. Our work is mainly focused on clicks, however, extension to other targets or multi-task learning is straight-forward.

\subsection{Overview of the Evaluator-Generator Framework}

We use $f_{\theta}(u, \textbf{a})$ to represent the predicted overall utility from the Evaluator, and $p(\textbf{a}|u, \textbf{c}) = \Pi_{\eta}(\textbf{a}; u, \textbf{c})$ to represent the Generator. We want $f_{\theta}$ to approximate the ground truth $r(u, \textbf{a})$, and we use $\Pi_{\eta}$ to plan or search for better score. The target of the Evaluation-Generation framework is to solve equation.~\ref{Eq:Evaluator-Generator}. 

\begin{equation}
\begin{split}
\label{Eq:Evaluator-Generator}
&argmax_{\textbf{a} \in \{\textbf{a}_1,...,\textbf{a}_n\}} f_{\theta}(u,\textbf{a})\text{,}\\
&\quad \text{with } f_{\theta}(u,\textbf{a}) \rightarrow \E[r(u, \textbf{a})] \text{ and } \textbf{a}_1, ..., \textbf{a}_n \sim \Pi_{\eta}
\end{split}
\end{equation}

\begin{figure*}
\centering
\includegraphics[width=10cm]{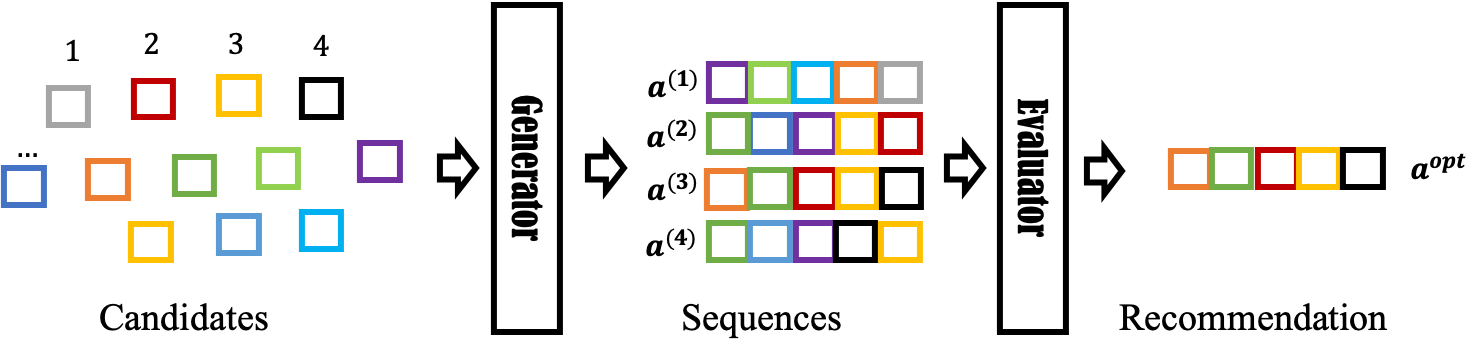}
\caption{A sketch of the Evaluation-Generation framework. The Generator generates potential high-quality sequences with heuristic searching or reinforcement learning, from which the Evaluator further select the superior one.}
\label{fig:EvaluatorGenerator}
\end{figure*}

We use the formulation $X_{\theta} \rightarrow Y$ here to represent that "Approximate $Y$ with model $X_{\theta}$ with respect to parameters $\theta$". Typically we reduce the mean square error (MSE) between $X_{\theta}$ and $Y$, or maximize the log likelihood. A sketch of the framework is shown in fig.~\ref{fig:EvaluatorGenerator}. On the one hand, we require the Evaluator to consider the sequence as a whole in order to fully capture position biases, diversity and other correlations etc before. We optimize the parameter $\theta$ by supervised learning to minimize the approximiation error. On the other hand, we require the Generator to effectively plan the sequences that achieves higher overall utility. We argue that the solution of equation.~\ref{Eq:Evaluator-Generator} can effectively approximate the solution of equation.~\ref{Eq:SRSTarget} by carefully selecting the model architecture and learning metrics. In the following part, we explain the model architectures and learning metrics for the Evaluator and Generator respectively.

\subsection{The Evaluators}

The Evaluator $f_{\theta}(u,\textbf{a})$ encodes the whole sequence and approximates the overall utility. We argue that $f_{\theta}$ should have the following characteristics: $f_{\theta}$ needs to be sensitive to the order of the sequences, such that relation with position can be properly considered; $f_{\theta}$ should add as little artificial hypothesis as possible in order to account for all possible correlations. We propose to use sequence encoding structures, such as recursive neural structure and self-attention layers to enable sufficient interaction among the displayed items. We use MSE loss $L(\theta; f_{\theta})$ for the optimization of $\theta$, written as equation.~\ref{Eq:Evaluator-SingleTarget}.

\begin{equation}
\label{Eq:Evaluator-SingleTarget}
\text{minimize } L(\theta; f_{\theta}) = \E_{u, \textbf{a}}[(f_{\theta}(u, \textbf{a}) - r(u, \textbf{a}))^2]
\end{equation}

While it is possible to directly predict the overall utility with $f_{\theta}(u,\textbf{a})$, the item-wise feedback improves the performance by providing finer supervision. Take the user click as example, we require the evaluator to predict $\E[\textbf{r}(u,\textbf{a})]$ rather than $\E[r(u, \textbf{a})]$. Thus we set the learning target as $\textbf{f}_{\theta} = [f_{\theta, 1}, ..., f_{\theta, K}]$, and $f_{\theta} = \sum_j f_{\theta, j}$, which gives the target loss function of equation.~\ref{Eq:Evaluator-MultiTarget}

\begin{equation}
\label{Eq:Evaluator-MultiTarget}
\text{minimize } L(\theta; \textbf{f}_{\theta}) = \E_{u, \textbf{a}}[\sum_{j=1}^{K}(f_{\theta,j}(u, \textbf{a}) - r_j(u, \textbf{a}))^2]
\end{equation}

Notice that until now we have made no proposition on the specification of the model architectures. There are whole bunches of neural network models proposed before to tackle different aspects of RS, including the well known Youtube RS(\cite{covington2016deep}), Google deep-and-wide RS(\cite{cheng2016wide}). The model structure is highly coupled with the problem and the features. As we focus mainly on the intra-list correlation, in order to make the comparison easier, we propose several relatively simple, but representational models in the following part. We can surely add more complexity to those structures to account for more features and factors (such as dynamic user interest), but those are not refered in this paper. 

We use $\boldsymbol{\phi}(u)$ and $\boldsymbol{\phi}(c_k)$ to denote the feature descriptor of user $u$ and item $c_k$ respectively, and $\boldsymbol{\phi}(j)$ to represent the position embedding. In our experiments, the dimension of the feature descriptor is $24$ for $u$, $32$ for $c_k$ and the embedding size is $8$ for position $j$. The feature descriptor of the user and the item is a mixture of the embedding vector and other dense features. A sketch of the model can be found in fig.~\ref{fig:ModelArch}. Here we use the expression ``Dense'' to represent a linear mapping function with bias and activation function.

\subsubsection{Multi Layer Perceptrons(MLP)}
The classic RS typically do not consider intra-list correlations at all, i.e. each item is evaluated independently based on the user information $\boldsymbol{\phi}(u)$, item information $\boldsymbol{\phi}(c_{a_j})$, and position information $\boldsymbol{\phi}(j)$ for considering the position bias. Our baseline model uses Multi-Layer Perceptrons(MLP) with the concatenation of the above three vectors as input(fig.~\ref{fig:ModelArch:sub1}).

\subsubsection{Gated Recurrent Neural Network(GRNN)}
We further propose to use GRNN(\cite{Graves2013Speech}) to encode the preceding sequence $\textbf{a}_j^-$ in order to capture the interactions between $\textbf{a}_j^-$ and $c_{a_j}$, which is followed by two layer MLP(fig.~\ref{fig:ModelArch:sub2}). Though applying single-direction recursive structures to inter-list correlations in RS has been reported before(\cite{hidasi2015session}), we report the first validation of GRNN in intra-list correlations. Compared with other diversified ranking algorithms, GRNN has the advantage that no additional assumption is introduced in the formulation of the correlation. However, GRNN has also deficiencies by presuming that $f_{\theta, j}$ is independent of $\textbf{a}_j^+$. It assumes that the user exams all recommendations in top-bottom manner only, which is also the assumption of the cascade click model(\cite{chuklin2015click}). However, other studies have revealed that $\textbf{a}_j^+$ does have impact in the overall performance of the $j$th position, including more sophisticated click models(\cite{wang2015incorporating}) and mouse tracking studies(\cite{diaz2013robust}). Motivated by this, we propose to use Bi-GRNN(\cite{ma2016end}) and self-attention structures.

\subsubsection{Bi-Directional Gated Recurrent Neural Network}
To further improve the precision and take the following exposure $\textbf{a}_j^+$ into account, we apply an second GRNN in reversed direction in addition to GRNN((\cite{ma2016end})), followed by similar MLP structures(fig.~\ref{fig:ModelArch:sub3})

\subsubsection{Transformer} 
\begin{figure*}
  \centering
  \begin{subfigure}{.296875\linewidth}
  		\centering
  		\includegraphics[width=\linewidth]{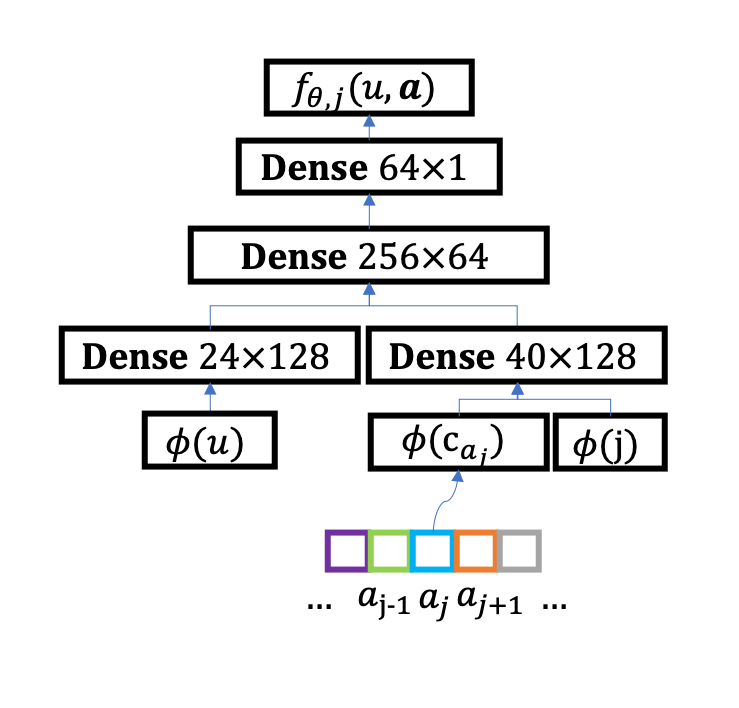}
  		\subcaption{}
  		\label{fig:ModelArch:sub1}
	\end{subfigure}
  \begin{subfigure}{.296875\linewidth}
  		\centering
  		\includegraphics[width=\linewidth]{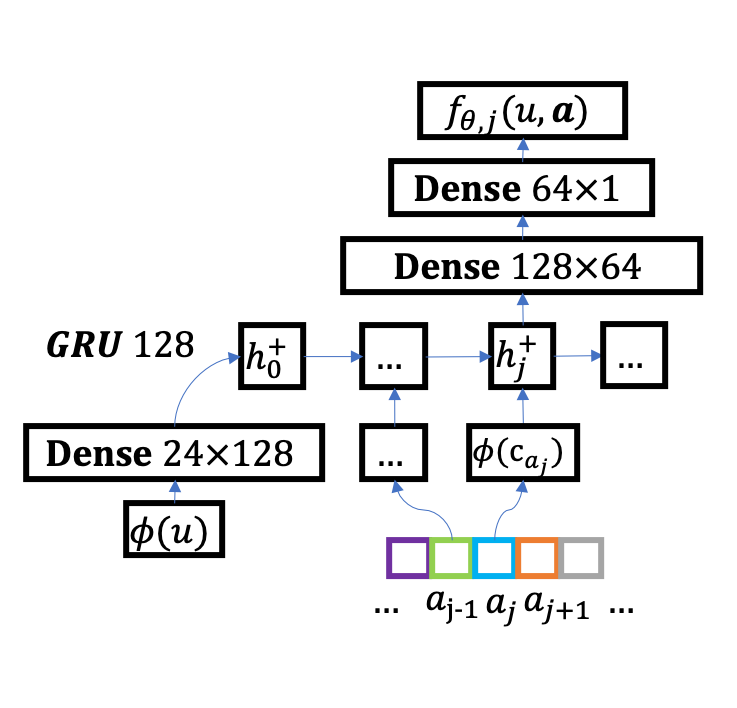}
  		\subcaption{}
  		\label{fig:ModelArch:sub2}
	\end{subfigure}
  \begin{subfigure}{.35625\linewidth}
  		\centering
  		\includegraphics[width=\linewidth]{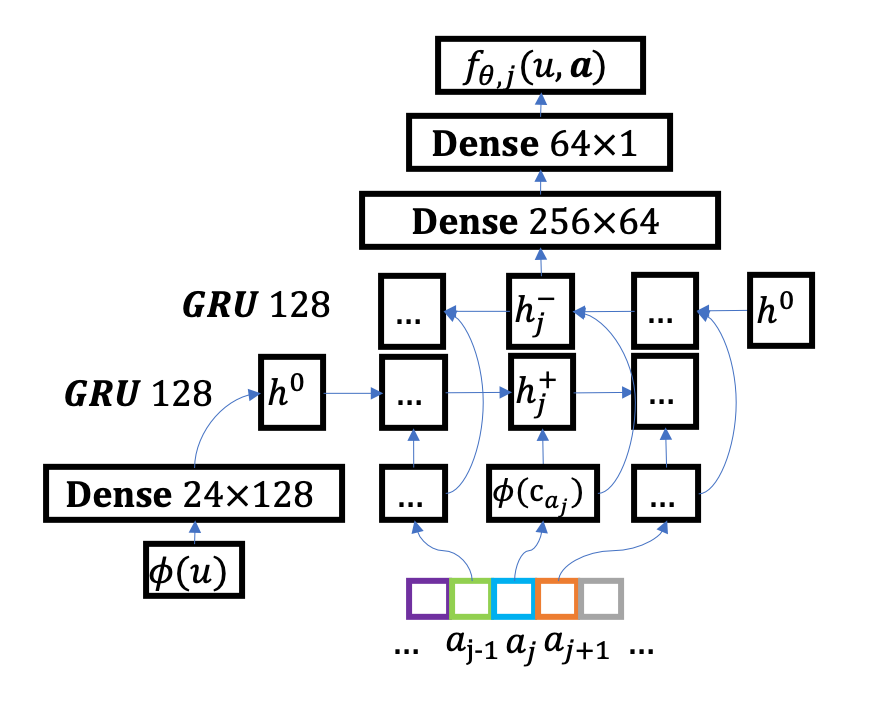}
  		\subcaption{}
  		\label{fig:ModelArch:sub3}
	\end{subfigure}
  \begin{subfigure}{.35625\linewidth}
  		\centering
  		\includegraphics[width=\linewidth]{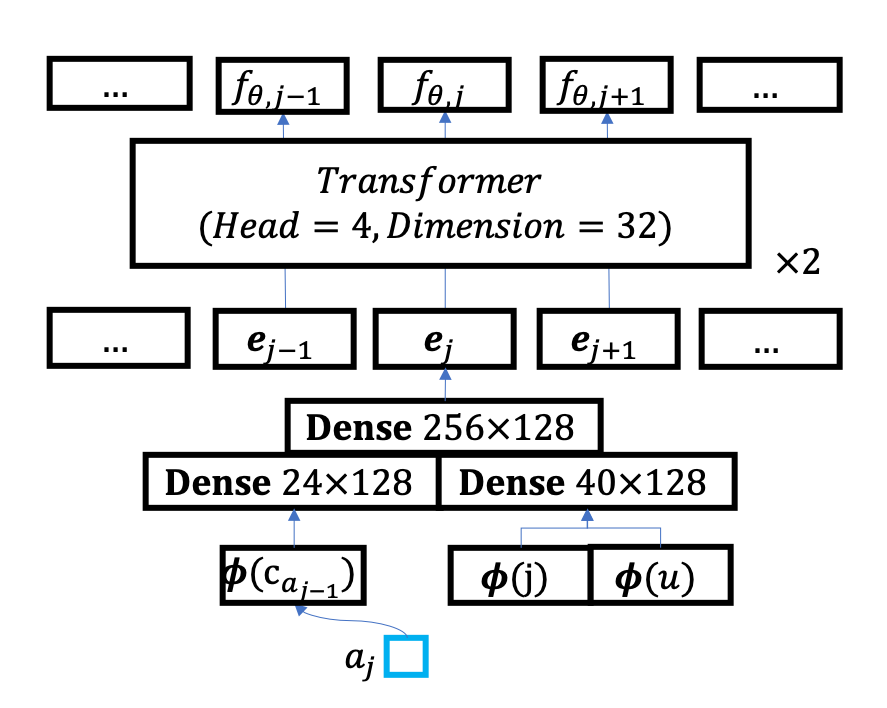}
  		\subcaption{}
  		\label{fig:ModelArch:sub4}
	\end{subfigure}
  \begin{subfigure}{.59375\linewidth}
      \centering
      \includegraphics[width=\linewidth]{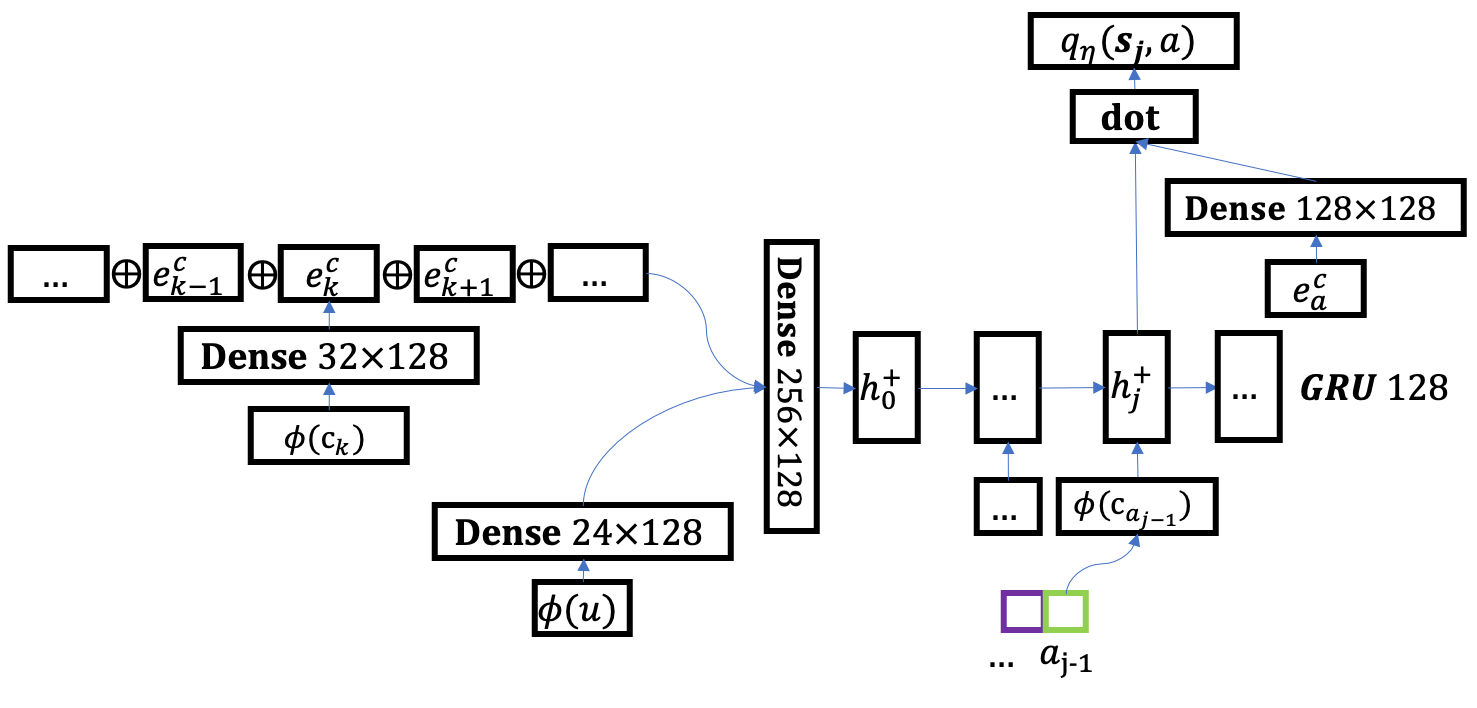}
      \subcaption{}
      \label{fig:ModelArch:sub5}
  \end{subfigure}
	\caption{Sketches of the proposed model architectures. (a). Multi-Perceptron Layers (MLP) as the Evaluator or the Generator; (b). Gated Recurrent Neural Network(GRNN) as the Evaluator or the Generator;
	(c). Bi-Directional Gated Recurrent Neural Network(Bi-GRNN) as the Evaluator;
	(d). Transformer as the Evaluator; (e). SetToSeq as the Generator}
  \label{fig:ModelArch}
\end{figure*}

Initially proposed for neural machine translation, Transformer(\cite{vaswani2017attention}) has achieved great successes as a sequence encoder. Self attention structure has been found to effectively capture the interactions intra-squence. We firstly concatenate the user descriptor with item representation and position embedding, and then we apply a 2-layer Transformer to predict the probability of click in each position(fig.~\ref{fig:ModelArch:sub4}).

\subsubsection{Other Works}.

While many previous works impose strong hypothesis on the formulations, Deep DPP (\cite{wilhelm2018practical}) has effectively combined the essence of DPP and the representation power of Deep Learning to push the intra-list modeling to new frontier. DPP approximate the utility of the sequence $\textbf{a}$ with $f_{\theta}(u, \textbf{a}) \propto |\textbf{K}_{\theta}(u, \textbf{a})|$, where $\textbf{K}_{\theta} = \{K_{\theta, ij}\}$ is a kernel matrix explicitly representing the correlation between each pair of position $i, j$, with $|\text{ }|$ representing the determinant of the matrix. Given the clicked subset $\textbf{a}_c = \{a_j | j \in [1,K], r_j = 1\}$, DPP tries to maximize $P_{\theta}(\textbf{a}_c)$, defined in equation.~\ref{Eq:DPP}

\begin{equation}
\label{Eq:DPP}
\text{maximize } P(\textbf{a}_c) = \frac{|\textbf{K}_{\theta}(u, \textbf{a}_c)|}{|\textbf{K}_{\theta}(u, \textbf{a}) + \textbf{I}|}.
\end{equation}

As proposed by Wilhelm et al, a neural mapping function can be used to map $u, \textbf{a}$ to the kernel matrix(\cite{wilhelm2018practical}). In this paper we use the model architecture shown in equation.~\ref{Eq:DPPModel}, where $D_{i,j}$ are the Jaccard distances between item descriptor $c_i$ and item $c_j$, and $\alpha$ and $\sigma$ are hyper-parameters.

\begin{equation}
\begin{split}
& e(u,k) = \textbf{Dense}([\boldsymbol{\phi}(u), \boldsymbol{\phi}(c_k)]), \\
& K_{\theta,ij}(u,\textbf{a}) = \alpha e(u,a_i) e(u,a_j) exp(-\frac{-D_{a_i, a_j}}{2\sigma^2}),
\end{split}
\label{Eq:DPPModel}
\end{equation}

\subsection{The Generators}

There exists various paradigm to generate an ordered sequence, such as deconvolution neural network(\cite{zhao2018deep}) and sequential generations. In our case we choose the latter one, where the probability of generating the sequence $\textbf{a}$, written as $\Pi_{\eta}(\textbf{a}; u, \textbf{c})$, is the product of sequential decisions $\pi_{\eta}$ shown in Eq.~\ref{Eq: Sequential Decision}.

\begin{equation}
\label{Eq: Sequential Decision}
\Pi_{\eta}(\textbf{a}; u,C) = \prod_{j=1}^K \pi_{\eta}(a_j; \textbf{s}_j)
\end{equation}

Here we denote the \emph{state} as $\textbf{s}_j = [u, \textbf{a}_j^-, \textbf{c}]$, the choice of the item index $a_j$ is the \emph{action} at step $j$. We use heuristic values to guide the policy $\pi_{\eta}(a_j; \textbf{s}_j)$, which we also call \emph{priority score}, denoted as $q_{\eta}(\textbf{s}_j, a)$. It represents the \emph{Value} of choosing action $a$ at the state $\textbf{s}_j$.  With the priority score we apply either greedy policy (equation.~\ref{Eq: Greedy_Geneartor}) or categorical sampling policy(equation.~\ref{Eq: Categorical_Sampling}, with $\tau$ being the hyperparameter of temperature).

\begin{equation}
\label{Eq: Greedy_Geneartor}
\pi_{\eta}(a; \textbf{s}_j) = 
\begin{cases}
1.0&\text{if $a = argmax_{a'} q_{\eta}(\textbf{s}_j, a')$} \\
0.0&\text{else}
\end{cases}
\end{equation}

\begin{equation}
\begin{split}
\label{Eq: Categorical_Sampling}
\pi_{\eta}(a; \textbf{s}_j) = \frac{exp(q_{\eta}(\textbf{s}_j, a)/\tau)}{\sum_{k=1}^N exp(q_{\eta}(\textbf{s}_j, k)/\tau)} 
\end{split}
\end{equation}

In order to effectively search for sequences that satisfies the target of equation.~\ref{Eq:Evaluator-Generator}, we can use supervised learning or reinforcement learning for the priority score $q_{\eta}$. In the following part we introduce different learning metrics and model structures for $q_{\eta}$.

\subsubsection{Supervised Learning}

Predicting the utility $\textbf{r}(u, \textbf{a})$ directly by learning from the user feedback are relatively simple and straightforward for $q_{\eta}$, which means we use the immediate feedback to approximate the priority score. Here the target function is shown in equation.~\ref{Eq:Generator-SL}. We use only $u$ and $\textbf{a}_j^-$ as the representation of state $\text{s}_j$, but we drop the information of $\textbf{c}$, as it is nearly of no help if we want to approximate the \emph{immediate} reward only. The MLP(which is similar to fig.~\ref{fig:ModelArch:sub1}) and GRNN(fig.~\ref{fig:ModelArch:sub2}) are reasonable choices for $q_{\eta}$ here, which do not rely on $\textbf{a}_j^+$. 

\begin{equation}
\begin{split}
\label{Eq:Generator-SL}
&\text{minimize } L(\eta; \textbf{q}_{\eta}) = \E_{u, \textbf{a}}[\sum_j(q_{\eta}(\textbf{s}_j, a_j) - r_j(u, \textbf{a}))^2]
\end{split}
\end{equation}

\subsubsection{Reinforcement Learning}

Applying RL enables the the sequential decision to pursue long term reward in each step. We argue that the choice of model architecture for RL needs to account for candidates $\textbf{c}$ in order to optimize the long term rewards. Here we propose a neural structure called the set to sequence decoder(SetToSeq for short).

SetToSeq borrows ideas from pointer-net(\cite{vinyals2015pointer}), which has been further combined with RL to solve universal combinatorial optimization(CO) problems(\cite{bello2016neural}). The main difference between SetToSeq and pointer-net lies in that exchangeable operations(we use summation) are applied to keep the model order insensitive to the candidate set $\textbf{c}$ in SetToSeq, while GRNN(\cite{Graves2013Speech}) or LSTM(\cite{hochreiter1997long}) is used to encode the candidates in pointer-net. Besides, we also inject the user and context information $\boldsymbol{\phi}(u)$ in the generative layer. The detailed model structure can be found in fig.~\ref{fig:ModelArch:sub5}.




Q Learning (\cite{mnih2013playing}) with replay memory is an efficient and widely used off-policy RL algorithm, where the value function $q_{\eta}(a, \textbf{s}_j))$ is to approximate the long term reward starting from the position $j$. As we are facing a finite horizon problem with static horizon($K$), we use $\gamma = 1.0$ as the decay factor. The temporal difference error is written as equation.~\ref{Eq:TemporalDifference}, where $\eta'$ is a delayed copy of parameter $\eta$, typically known as target network.

\begin{equation}
\label{Eq:TemporalDifference}
\begin{split}
&L_{TD}(\eta, q_{\eta}) = \E_{u, \textbf{a}}[(\max_a\{q_{\eta'}(a, \textbf{s}_j)\} + f_{\theta, j}(u, \textbf{a}) - q_{\eta}(a_j, \textbf{s}_j))^2]
\end{split}
\end{equation}

Here we use the \emph{simulated feedback} $f_{\theta, j}(u, \textbf{a})$ instead of the the \emph{real feedback} $r_j$ as the reward, which is directly derived from equation.~\ref{Eq:Evaluator-Generator}. However, it is also possible to use the real feedback $r_j$. We put some comments on the two kinds of feedback. Learning from simulated feedback takes risks as if the Generator "attacks" the Evaluator. In case discrepancies exist between the Evaluator and the realistic environment, the Generator would deviate from the real target. On the other hand, learning from real feedback in a realistic system typically requires quite heavy engineering work, or else it would result in off-policy learning with static data. Studies have revealed that off-policy learning with static data sometimes is not guaranteed to work well enough(\cite{fujimoto2018off}).

\section{Experiments}

In our experiments, we use 100 million lists from user-system interaction records for training, and 1 million lists for testing, which were collected from \textbf{B}aidu \text{A}pp \textbf{N}ews \textbf{F}eed \textbf{S}ystem (\textbf{BANFS}), one of the largest RS in China. BANFS has over hundreds of millions of daily active users. A sequence of $10\sim50$ items are refreshed corresponding to the user requirement. In our experiment, to reduce the cost of the experiment, our offline dataset \footnote{Our code is released at https://github.com/LihangLiu/Generator-Evaluator, which is based on PaddlePaddle. Desensitized dataset will come soon.} contains only a subset of the features, including the user id, item id, item category and layout (the appearance of the item). In our experiment settings, we focus mainly on the clicks of the recommendation, again extension to multi-targets are straight-forward but not mentioned here.

\subsection{Evaluation Metrics}
Traditional IR evaluation metrics(such as NDCG, MAP, ERR) are not suitable as those are based on static data. Previous work toward evaluating combinatorial RS include Yue et al. using an artificial simulator (\cite{yue2011linear}), others using online experiments for evaluation (\cite{wilhelm2018practical}). Also there are some counterfactual evaluation tricks (\cite{dudik2011doubly}, \cite{li2015counterfactual}), but applying those metrics to realistic RS with over millions of candidate items and users is often intractable.

In this work, we evaluate our ideas from the following criterions.
\begin{itemize}
\item Firstly, the precision of the Evaluators are evaluated by three metrics with realistic user feedback. Area Under Curve(AUC) of the ROC curve is used to evaluate the precision of predicting the utility of each position in the sequence. \textbf{SeqRMSE} and \textbf{SeqCorr} are used to evaluate the precision of predicting the overall utility. \textbf{S}equence \textbf{R}oot \textbf{M}ean \textbf{S}quare \textbf{E}rror (\textbf{SeqRMSE}) is defined as

\begin{equation}
SeqRMSE = \sqrt{\E_{u, \textbf{a}}[(f_{\theta}(u,\textbf{a}) - r(u, \textbf{a}))^2]} .
\label{Eq:RMSES}
\end{equation}

Since some methods, such as DPP, do not predict $\E[r(u,\textbf{a})]$ explicitly, we also evaluate the correlation between the overall utility $r(u,\textbf{a})$ and the evaluation score $f_{\theta}(u,\textbf{a})$(\textbf{SeqCorr} for short), which is defined in Eq.~\ref{Eq:Corr}.

\begin{equation}
\begin{split}
& SeqCorr = \frac{Cov(f_{\theta}(u,\textbf{a}), r(u,\textbf{a}))}{\sqrt{Var(f_{\theta}(u,\textbf{a})) \cdot Var(r(u,\textbf{a}))}} .\\
\end{split}
\label{Eq:Corr}
\end{equation}

\item Secondly, we compare different Generators by regarding the Evaluator itself as an simulator(or environment). Previous work has proposed using simulators to evaluate combinatorial recommendation(\cite{yue2011linear}). Building simulators to evaluate RL recommender systems offline has also been reported(\cite{shi2018virtual}). In our case, we argue that the Evaluator can work as both an \textbf{Selector} online and a natural \textbf{Simulator} offline. We do not only provide the comparison of different Generators under the proposed simulator, but we also demonstrate the validity of the simulator itself by directly comparing the Evaluator to the online environments.

\item Finally, we publish the result to compare different ranking frameworks in online A/B tests. As we are more cautious toward online experiments, we did not carry out the experiments on all possible ranking frameworks. It is worth noticing that our online experiments uses larger feature set and datasets to achieve better performance, thus the performance of the online and offline experiments are not totally comparable. 

\end{itemize}

\subsection{Results on The Evaluators}

To evaluate the precision of the Evaluator models, we use AUC, SeqRMSE and SeqCorr as the criteria for comparison. We compare the MLP, GRNN, Bi-GRNN and the Transformers, the architectures of which are shown from fig.~\ref{fig:ModelArch:sub1} to fig.~\ref{fig:ModelArch:sub4}. We use adam optimizer(\cite{Kingma15Adam}) with the batch size of $768$ and the hyper-parameters of $\text{learning rate} = 1.0e-3, \beta_1 = 0.9, \beta_2 = 0.999$. For each model the optimizer goes through the training data for $20$ epochs, and the last epoch is used. 

By concluding from Tab.~\ref{Tab: Evaluator-AUC}, we can see that $\textbf{a}_j^-$ and $\textbf{a}_j^+$ do make impact on the click of the $j$-th position, as the Bi-GRNN and the Transformer outperforms the other models in all three evaluation criteria. The performance of DPP is below the baseline, which is mainly caused by missing the position bias, while in BANFS the clicks are severely influenced by its position

\begin{table}[h!]
\begin{center} 
  \begin{tabular}{m{3.2cm}<{\centering}|m{1.2cm}<{\centering}m{1.2cm}<{\centering}m{1.2cm}<{\centering}}
    \hline
    \textbf{Algorithm} & AUC & SeqRMSE & SeqCorr \\
    \hline \hline
    MLP & 0.7706 & 0.4478 & 0.4733 \\
    GRNN & 0.7754 & 0.4443 & 0.4853 \\
    Bi-GRNN & 0.7798 & 0.4420 & 0.4936 \\
    Transformer & 0.7794 & 0.4436 & 0.4906 \\
    Deep DPP & - & - & 0.3810 \\
    \hline \hline
  \end{tabular}
  \caption{Offline comparison of different Evaluators}
  \label{Tab: Evaluator-AUC}
\end{center}
\vspace{-0.6cm}
\end{table}

\begin{figure*}
  	\centering
    \begin{subfigure}{.33\linewidth}
  		\centering
  		\includegraphics[width=\linewidth]{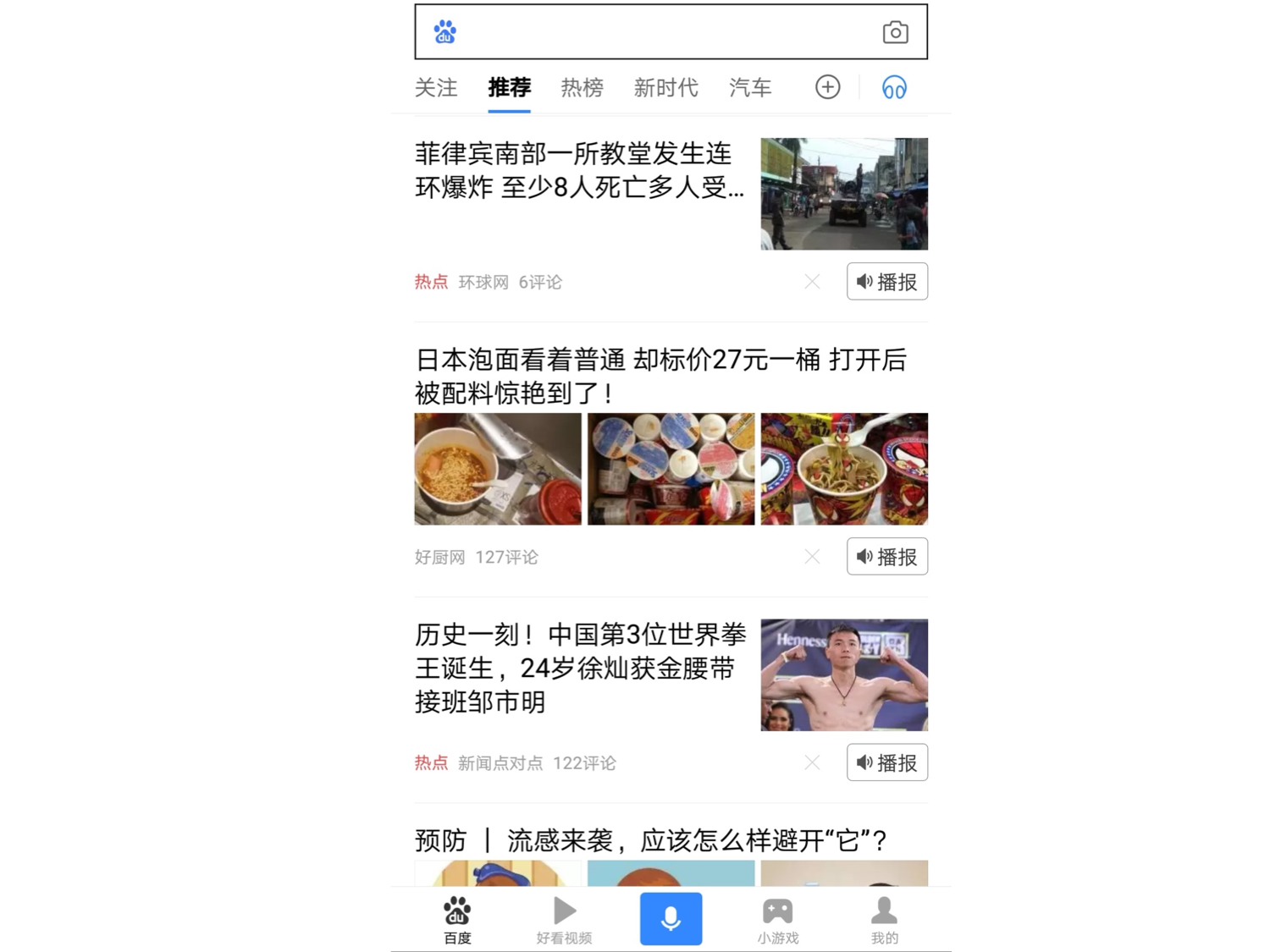}
  		\subcaption{}
  		\label{fig:Attention:sub1}
	\end{subfigure}
    \begin{subfigure}{.33\linewidth}
  		\centering
  		\includegraphics[width=\linewidth]{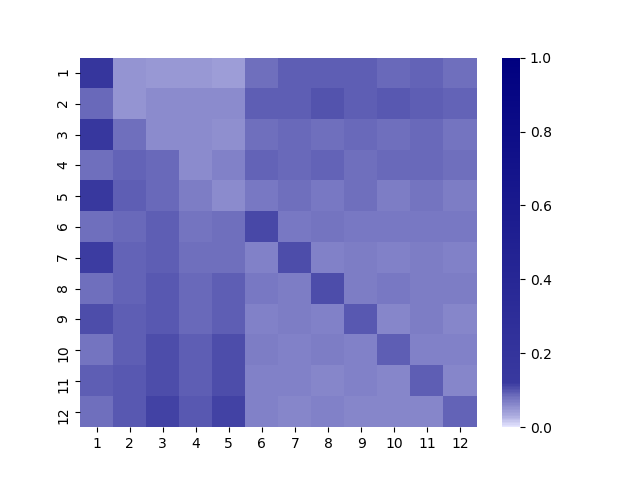}
  		\subcaption{}
  		\label{fig:Attention:sub2}
	\end{subfigure}
    \begin{subfigure}{.33\linewidth}
  		\centering
  		\includegraphics[width=\linewidth]{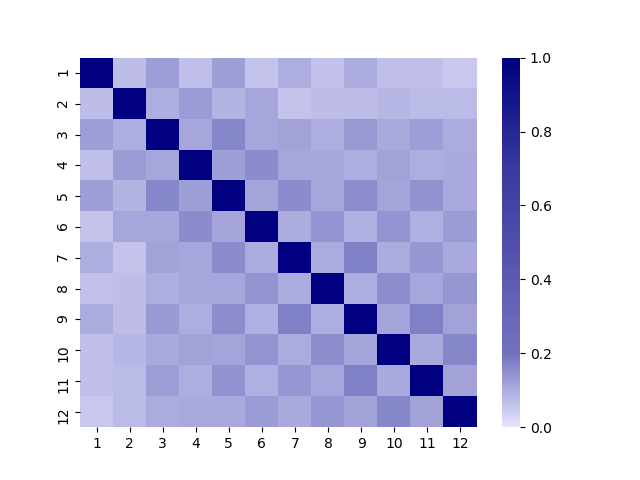}
  		\subcaption{}
  		\label{fig:Attention:sub3}
	\end{subfigure}
	\caption{(a) A snapshot of BANFS.  (b) Average attention weights between any two positions(x and y) over 6000 lists. (c) Correlation of realistic user clicks between any two positions, statistics from the same lists in user log.}
    \label{fig:Attention}
    
\end{figure*}

To visualize the intra-list correlation in BANFS, we plot the heat-map of the average attention strength on the self-attention layer of first layer in the Transformer(Fig.~\ref{fig:Attention:sub2}). We mention several phenomenons that are coherent with our instinct or investigation: Firstly, each item is more dependent on its previous items, especially those on the top of the list. The BANFS distributes 10 $\sim$ 50 items as a sequence at each time, but only 2 $\sim$ 4 items can be displayed in a screen(Fig.~\ref{fig:Attention:sub1}). A user needs to slide downward to examine the whole list. Thus, whether the items on the top of the sequence attracts the user has an impact on the probability that the items lie below are examined, and thus clicked. Secondly, the attention between adjacent positions $j$ and $j+1$ is not as strong as that between $j$ and $j+2$, which makes the heat-map interweaving(like chessboard). To better study this phenomenon, we further plot the realistic correlation of clicks between each position pair. Fig.~\ref{fig:Attention:sub3} shows that the correlation of user clicks is interweaving: the adjacent positions is less likely to be clicked together, but $j$ and $j+2$ is more likely to be clicked together. This is in consistency with the \textbf{Serpentining} phenomenon that was mentioned in \cite{zhao2017toward}. This phenomenon has further shown that the intra-list correlation is much more complicated than many position bias hypothesis or unordered set-wise hypothesis previously proposed. 

\subsection{Results on the Generators}

To evaluate the Generators, we randomly sample 1 million candidate sets from the user interaction logs. Those are regarded as pools of candidate $\textbf{c}$. We begin a simulation where we randomly sample an user $u$, a candidate set $\textbf{c}$ and length of the final list $K$ in each step. The length $K$ follows the real distribution of sequence length online, which varies between 10 and 50. We sample the candidate set such that $N=2K$. Then, the Generator is required  to generate one or multiple sequences of length $K$, and we use the Evaluator to pick the sequence with highest overall score (if only 1 list is generated, such as greedy picking, then there are no need of using Evaluators as a selector). We compare the statistical evaluation score $\E_{u, \textbf{a} \sim \Pi_{\eta}}[f_{\theta}(u, \textbf{a})]$ on the generated lists. For the training settings, in SL we keep the hyper-parameters the same as that of the Evaluator; in RL we use a replay memory of size $768K$, and we keep the training batch numbers and learning rates the same as SL.

We choose three different Evaluators as the simulator, including GRNN, Bi-GRNN and Transformer. We use three different model architectures for the Generators(MLP, GRNN and SetToSeq), which is combined with different learning metrics(SL and RL) and policies(Greedy and Sampling). Notice that in case we use GRNN + SL as the Evaluator and the Generator at the same time, the sampling policy can also be replaced with beam search. We show the results in Tab.~\ref{Tab: Generator}. Several remarks can be made, the comparison among different model architectures shows the importance of incorporating contexts $\textbf{a}^-_j$ and candidates $\textbf{c}$.  The comparison of greedy policy and sampling policy shows that there is indeed non-negligible gap between local and global optimum. The RL group outperforms SL group under comparable policy, showing that RL can be used to improve the efficiency of searching for global optimum, or to reduce the number of the sampled trajectories which is required to achieve comparable performance.

\begin{table*}[h!]
\begin{center} 
  \begin{tabular}{c|c|c|c|c|c|c}
    \hline 
    \multirow{3}{*}{\textbf{Generator}} & \multicolumn{6}{c}{\textbf{Evaluator}} \\
    \cline{2-7}
    & \multicolumn{3}{c|}{\textbf{Bi-GRNN}} & \multicolumn{3}{c}{\textbf{Transformer}} \\
    \cline{2-7}
    & Greedy & Sampling($n=20$) & Sampling($n=40$) & Greedy & Sampling($n=20$) & Sampling($n=40$) \\
    \hline \hline
    \textbf{MLP + SL} & 1.3538 & 1.7030 & 1.7540 & 1.3615 & 1.6290 & 1.6779 \\
    \textbf{GRNN + SL} & 1.6652 & 1.8852 & 1.9226 & 1.5522 & 1.7853 & 1.8267 \\
    \textbf{GRNN + RL(Simulated Data)} & 1.7552 & 1.9399 & 1.9683 & 1.8296 & 2.0239 & 2.0550 \\
    \textbf{SetToSeq + RL(Simulated Data)} & 1.9174 & 2.0313 & 2.0606 & 1.9751 & 2.1000 & 2.1348 \\
    \textbf{SetToSeq + RL(Real Data)} & 1.3449 & 1.5851 & 1.6259 & - & - & - \\
    \hline
  \end{tabular}
  \caption{Offline comparison of different Generators by using the Evaluators for simulation.}
  \label{Tab: Generator}
\end{center}
\vspace{-0.5cm}
\end{table*}

\begin{figure*}
  	\centering
    \begin{subfigure}{.49\linewidth}
  		\centering
  		\includegraphics[width=\linewidth]{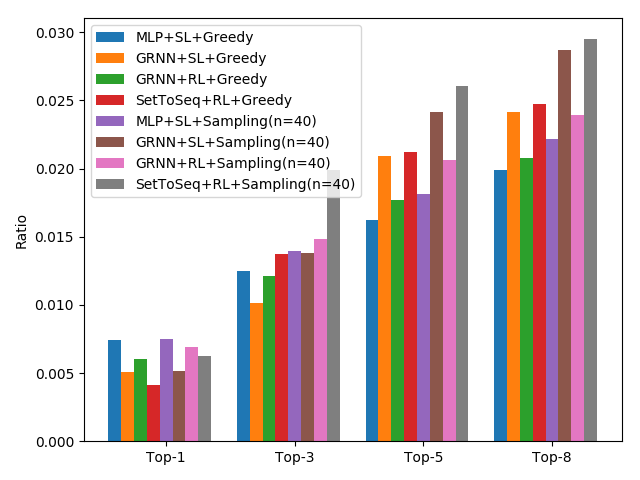}
  		\subcaption{}
  		\label{fig:Histogram:sub1}
	\end{subfigure}
    \begin{subfigure}{.49\linewidth}
  		\centering
  		\includegraphics[width=\linewidth]{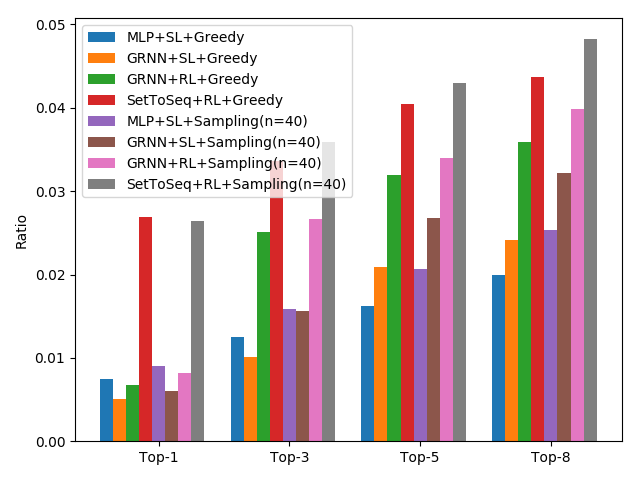}
  		\subcaption{}
  		\label{fig:Histogram:sub2}
	\end{subfigure}
	\caption{The ratio of Top-K patterns to all generated patterns($P_{l_{s_j}, l_{s_{j+1}}, l_{s_{j+2}}}$) with different Evaluation-Generation frameworks. (a) Bi-GRNN as the Evaluator. (b) Transformer as the Evaluator}
    \label{fig:Histogram}
    
\end{figure*}

To illustrate that the proposed framework indeed yields better item combinations, we did some inspection in the generated lists. The combination of item ids or high dimensional features are far too sparse to give any insightful results, thus we focus on the layouts of the short local segments. The layout marks the visual content arrangement within each item when shown to the users, as shown in Fig. \ref{fig:Attention:sub1}. More concretely, we denote the layout of the $i$-th item as $l_i \in [1, M]$, where $M$ is the size of all distinct layouts. Then we consider the layouts of a segment of three consecutive items in position $(j, j+1, j+2)$ as the local pattern $P_{l_{s_j}, l_{s_{j+1}}, l_{s_{j+2}}}$ that we want to look into. In BANFS, there are $M=6$ types of layouts, e.g. "text-only", "one-image", "three-images", etc. Thus there are $6^3 = 216$ distinct layout patterns in total. For a recommendation list of length $K$, we count all possible $K-2$ segments. The procedure of analyzing local patterns works as follows: Firstly, the average sum of click $\E[r(P_{l_{s_j}, l_{s_{j+1}}, l_{s_{j+2}}})]$ of the segment pattern can be counted from the user log. Under the assumption that the higher click rate means better quality of the pattern, we use $\E[r(P_{l_i, l_j, l_k})]$ to measure the quality of the local layout pattern $P_{l_i, l_j, l_k}$. So we regard the layout patterns in 216 possible patterns that rank top-N in the expected clicks as "good patterns". To evaluate our proposed framework, we calculate the ratio of the top-N pattern segments in the generated lists from different Generators. We list the distribution on Top-K patterns with different ranking frameworks in fig.~\ref{fig:Histogram}. The figure shows that the list with higher evaluation score indeed include more "good patterns", which also means the evaluation score is consistent with intuitive indicators. 

\subsection{Performance Online}

\textbf{Correlation between Evaluators and Online-Performance} The previous results show that the Evaluator is more correlated to the sum of clicks of a list. But, is the predicted sum of clicks related to the final performance? Is it appropriate to treat Evaluator as a simulator? We perform additional investiagtion on the correlation between the Evaluation score of lists $f_{\theta}(u, \textbf{a})$ and the performance of A/B test. Typically we judge whether a new policy is superior than the baseline, by the increase in some critical metrics, such as total user clicks, in A/B test. For two experiment groups with experiment ID $I_A$(experimental) and $I_B$(baseline), the \textbf{Relative Gain} in total clicks is defined as the relative increase of clicks compared with baseline. Thus we retrieve the logs of the past experiments, and we re-predict click of each sequences in the record by inferencing with our Evaluator model. We calculate the \textbf{Predicted Relative Gain} by

\begin{equation}
\begin{split}
    \label{Eq:Click_Gain}
   & \Delta = \frac{\sum_{i \in I_A} f_{\theta}(u,\textbf{a}_i) - \sum_{i \in I_B} f_{\theta}(u,\textbf{a}_i)}{\sum_{i \in I_B} f_{\theta}(u,\textbf{a}_i)}. \\
\end{split}
\end{equation}

We collect over 400 A/B testing experiments during 2018, including not only new ranking strategies with various policy, model and new features, but also human rules. Some of the new strategies are tested to be positive, others negative. We counted the correlation between the predicted relative gain and the statistical real relative gain. We use MLP and Bi-GRNN Evaluator for comparison. The correlation between MLP and online performance among the 400 experiments is $0.2282$, while the correlation betwen Bi-GRNN and real performance is as high as $0.9680$. It has proved to some extent that the Evaluator can evaluate a strategy before doing A/B test online and the confidence is relatively high, thus the simulation results by treating the Evaluator as the simulator are relatively confident.

\begin{table}[h!]
\begin{center} 
  \label{Tab: Correlation_Online}
  \begin{tabular}{m{4.0cm}<{\centering}|m{1.5cm}<{\centering}|m{1.5cm}<{\centering}}
    \hline
     & MLP & Bi-GRNN \\
    \hline \hline
    Correlation with Online Performance & 0.2282 & 0.9680 \\
    \hline
  \end{tabular}
  \caption{Correlation between the Evaluator predictions and the online A/B tests}
\end{center}
\vspace{-0.6cm}
\end{table}

\begin{table}[h!]

\end{table}

\textbf{Online A/B Test of Ranking Framework} In order to finally validate the proposed framework, we have conducted a series of online experiments on BANFS, we report the comparison of the following methods.

\begin{itemize}

\item \emph{MLP}: Generator only, MLP + SL (Greedy)

\item \emph{GRNN}: Generator only, GRNN + SL (Greedy)

\item \emph{Evaluator-Generator}: GRNN + SL + Sampling($n=20$) as the Generator, Bi-GRNN as the Selector. 

\item \emph{SetToSeq + RL + Simulated Data}: Training the SetToSeq model with Q-Learning and simulated feedback. The simulated feedback comes from interacting with the Evaluator of Bi-GRNN.

\item \emph{SetToSeq + RL + Real Data}: Training the SetToSeq model with Q-Learning and realistic clicks, in completely off-policy manner with finite log data.

\end{itemize}

\begin{table*}[h!]
\begin{center} 
  \begin{tabular}{m{8.0cm}<{\centering}|m{3.0cm}<{\centering}|m{3.0cm}<{\centering}}
    \hline
    Recommender Framework &  Relative Gain ($\%$) & Coverage of Categories($\%$) \\
    \hline \hline
    \textbf{GRNN} vs. \textbf{MLP} &  $+1.75$ & $+4.05$ \\
    \textbf{Evaluator-Generator} vs. \textbf{GRNN} & $+2.44$ & $+0.62$ \\
    \textbf{SetToSeq + RL + Simulated Data} vs. \textbf{Evaluator-Generator} & $-0.01$ & - \\
    \textbf{SetToSeq + RL + Real Data} vs. \textbf{Evaluator-Generator} & $-0.05$ & - \\
    \hline
  \end{tabular}
  \caption{Comparison of different solutions in online performance}
  \label{Tab:OnlineExperiments}
\end{center}
\end{table*}

Our evaluation result until now shows that Evaluator-Generator with softmax sampling Generator has state-of-the-art online performance. The RL group has shown comparable performance with Evaluator-Generator. Though we believe that RL should be more cost-efficient and more straightforward for solving the intra-list correlation, our experiments shows that the performance of RL occasionally generate unexpected bad patterns. We have also compared the Coverage(\cite{ziegler2005improving}) on distinct categories(There are 40+ categories in all). It is verified that the proposed framework indeed improves diversity of exposure even though diversity is never considered as a explicit target in our framework(tab.~\ref{Tab:OnlineExperiments}).

\section{Discussions}

In this paper, we propose a recommender framework by optimizing K-item in the result page as a whole. We propose the Evaluation-Generation framework to solve the combinatorial optimization problem. We show that compared with traditional diversified ranking algorithms, the proposed framework is capable of capturing various possible correlations as well as the position bias. In this section, we post some of our further comments in this framework and its possible future extensions.

\textbf{Robustness of RL}. Though Q-learning greatly outperformed the other learning metrics in off-line Evaluations, it is found that Q-learning is vulnerable to the noisy online policy. E.g., when some positions in the sequence are disturbed by the other mandatory interventions(which is normal in online system), the model can generate poor combinations. The robustness of RL in our case deserves further investigation.

\textbf{Exploration and Exploitation}. Exploration is important for interactive systems, as continuously greedy recommendation would end up in mediocre or outdated contents. We propose that RS should also explore different combination of items besides the item itself. Suppose that the model see only "good" combinations, the system would not be able to learn to avoid "bad" ones such as duplicated recommendations. Thus we keep a small fraction of PV for exploring different combinations randomly online.

\textbf{Synthesising Intra-list and Inter-list Relations}. Typical RS has both the features of intra-list correlation and inter-list evolution. However, building unified framework to address both factors in realistic RS remains challenging. We believe it's a promising direction toward the next generation RS.




\bibliographystyle{named}
\bibliography{SequenceBasedRecommendation}

\end{document}